\documentclass[dvips]{article}
\usepackage{icrc2011}
\usepackage{amsmath}
\usepackage{graphicx}

\title{Variability along the Blazar-Sequence - Hints for extragalactic Cosmic Rays?}

\newcommand{\etal}{\MakeLowercase{\textit{et al. }}} 
\newcommand{\sourceone}{\mbox{Mkn 501}}
\newcommand{\sourcetwo}{\mbox{1 ES 2344+51.4}}
\newcommand{\sourcethree}{\mbox{PKS 2155-30.4}}
\newcommand{\sourcefour}{\mbox{1 ES 1218+30.4}}
\newcommand{\sourcefive}{\mbox{3C 279}}
\newcommand{\sourcesix}{\mbox{3C 454.3}}
\shorttitle{M. Weidinger \etal Variability along the Blazar-Sequence}

\authors{Matthias Weidinger, Felix Spanier}
\afiliations{Institut f\"ur Theoretische Physik und Astrophysik, Universit\"at W\"urzburg}
\email{mweidinger@astro.uni-wuerzburg.de}

\abstract{The spectral energy distribution and variability of several blazars (\sourceone, \sourcetwo, \sourcethree, \sourcefour, \sourcesix) along the blazar sequence is investigated using a selfconsistent and timedependent lepto-hadronic hybrid emission model. The BL Lac objects in the list are successfully modelled with purely leptonic radiation processes (essentially Synchrotron Self-Compton emission), while the Flat Spectrum Radio Quasar requires highly relativistic hadrons to be present within the jet. Variability is exploited as well with our model to distinguish between Self-Compton and hadronic gamma radiation making use of their different signatures in lightcurves. As a consequence active galactic nuclei with $z > 0.5$ are the best candidates as sources of extragalactic consmic rays, since High-Peaked BL Lac objects do not seem to accelerate protons to energies above thermal. Furthermore the parameters found during the modelling of the objects agree very well with obervations of e.g. superluminal motion or typical variability timescales.}

\begin{document}
\maketitle

\section{Introduction}
Blazars have gained a lot of attraction in the past decades, mainly due to the discovery of their very high energy (VHE) emission up to $10^{27}~\text{Hz}$ with Air-Cherenkov Telescopes like H.E.S.S. or MAGIC. Ever since more and more blazars with a redshift up to $z = 0.536$ (\sourcefive~ with MAGIC \cite{magic279}) are discovered in the VHE making a systematic investigation possible. The spectral energy distribution of blazars shows two prominent humps, one in the optical to hard X-Ray band and one in gamma-ray energies. The emission is variable on short time scales across the whole spectrum with large amplitude changes. Fermi extended our view allowing a nearly all-sky survey between $20~\text{GeV}$ and $300~\text{GeV}$. Thus the availability of multiwavelength (MWL) data, even during outbursts of blazars, is rapidly increasing.\\
Blazars are phenomenologically divided into different subclasses based on their peak frequency forming the blazar sequence: Flat Spectrum Radio Quasars (FSRQs) like \sourcefive~as the most luminous objects having their first peak in the near infrared to optical regime. As the overall luminosity decreases in Low frequency and Intermediate frequency peaked BL Lac objects (LBLs and IBLs respectively) the first hump shifts to higher energies with high frequency peaked BL Lac objects (HBLs) achieving the highest photon energies up to $2~\text{TeV}$ showing the lowest luminosity among blazars \cite{ghisellini, urry}.\\
While HBLs are well explained by Synchrotron Self-Compton (SSC) models \cite{finke, weidinger02}, i.e. the first peak is due to synchrotron radiation of primary electrons and the second hump arises from Compton upscattering (IC) of the very same radiation by the primary particles, LBLs and FSRQs may not be explained the same model. Their hard VHE radiation and wide spread of the peaks are not consistent with IC scattering of synchrotron photons \cite{boettcher3C279}. The origin of the VHE emission in this subsample is still under passionate debate: either external radiation providing the target photons for IC upscattering within the jet \cite{boettcherEX}, or it is due to photo-hadronic cascades initiated by high energy protons combined with their synchrotron radiation \cite{boettcher3C279, mannheim93}.\\
In this paper we use a selfconsistent, timedependent hybrid model co-accelerating electrons and protons with Fermi-I and Fermi-II processes at the very same site while taking all the relevant loss processes into account to reproduce the emission of different blazars along the sequence with just one model by varying the parameters like the $p^+$ to $e^-$ ratio. Outbursts along the sequence can be exploited with the model as well, supporting e.g. the discrimination against EC enhanced models or the determination of sources accelerating protons and not only electrons to the highest energies.\\
Consequently FSRQs are the most promising candidates as sources of extragalactic ultra high energetic cosmic rays (UHECRs) among blazars. The UHECRs are believed to be extragalactic since we clearly see the GZK-Cutoff in the well known spectrum of CRs \cite{hires, auger} and due to the lack of a large-scale anisotropy in the spectrum towards the galactic disk. Of course these UHECR statements are strongly limited by statistics.

\section{The Model}
In this section we briefly introduce our model focusing on the additions compared to the more detailed description already available in \cite{weidinger01, weidinger02}. The geometry and general properties of the model stay untouched while the corresponding one dimensional Vlasov-equation in the diffusion approximation (see e.g. \cite{schlickeiser02}) is solved for electrons and protons separately in two spatially separated zones. The upstream spherical acceleration zone is nested in a bigger radiation zone. The kinetic equations in the acceleration zone therefore are
\begin{align}
 \label{acc}
	\partial_t n_i = &\partial_{\gamma} \left[( \beta_{s,i} \gamma^2 - t_{\text{acc,i}}^{-1}\gamma ) \cdot n_i\right]  +\nonumber \\& \partial_{\gamma} \left[ [(a+2)t_{\text{acc,i}}]^{-1}\gamma^2 \partial_{\gamma} n_i\right] + Q_{0,i} - \frac{n_i}{t_{\text{esc,i}}}
\end{align}
with $i$ being electrons (e) or protons (p) respectively. The synchrotron $\beta_s$ and $t_{\text{esc}}$ depend on the considered particles' mass as follows (CGS units)
\begin{align}
 \label{massdep}
 	\beta_{s,i} & = \frac{\sigma_{\text{T}}B^2}{m_ic} = \frac{8\pi}{3} \frac{q_i^4 B^2}{m_i^3c^6}~,~~t_{\text{esc,p}} = \frac{m_p}{m_e} t_{\text{esc,e}}~\text{.}
\end{align}
By setting $t_{\text{acc}}/t_{\text{esc}}$ as a parameter during the modeling the acceleration timescale will scale linearly with the mass as one would expect, i.e. $\gamma_{\text{max,i}}$ particles can reach scales with $m_i^2$, whereas the spectral index being independent since the acceleration mechanism itself stays the same, i.e.:
\begin{align}
 \label{gmaxratio}
	\gamma_{\text{max,i}} & = \frac{1}{\beta_{s,i}t_{\text{acc,i}}} ~ \Rightarrow ~ \frac{\gamma_{\text{max,p}}}{\gamma_{\text{max,e}}} = \left(\frac{m_p}{m_e}\right)^2~\text{.}
\end{align}
Note that the only new parameter introduced compared to the simple SSC case is the injection height $Q_{0,p}$ of protons, but due to the opacity to $p+\gamma$ processes this has a major effect on the radiation zone. The plasmaphysical properties like $K_{||}$, $v_{\text{shock}}$ or $v_{\text{Alfv\'en}}$ can still be calculated as described in \cite{weidinger01, weidinger02}.\\
Protons with non-thermal energies are above the threshold for pion production $p+\gamma \rightarrow p + n_0 \pi^0 + n_+ \pi^+ + n_-\pi^-$ within the jet which decay into electrons, positions and $\gamma$s (and neutrinos), the channels are
\begin{align}
 \label{pgammachain}
	\pi^{+} & \rightarrow \mu^{+} + \nu_{\mu} \rightarrow e^{+} + \nu_e + \nu_{\mu} \nonumber \\ \pi^{-} & \rightarrow \mu^{-} + \bar{\nu}_{\mu} \rightarrow e^{-} + \bar{\nu}_e + \bar{\nu}_{\mu} \nonumber \\ \pi^0 & \rightarrow \gamma + \gamma
\end{align}
This will result in the third contribution, besides proton synchrotron and IC from electrons, to the VHE peak in the SED of blazars when hybrid models are considered. We use the Monte-Carlo simulations of \cite{kelneraharonian} to determine the production rate of the stable electrons, positrons and $\gamma$s from the interactions of eq. \eqref{pgammachain}. We do not account for synchrotron losses of the unstable $\pi^{\pm}$, $\mu^{\pm}$, i.e. the photohadronic interactions are assumed to be instantaneous compared to the synchrotron loss timescale. The produced $\gamma$s have energies well above $511$ keV and will therefore result in pair cascades of $e^{\pm}$. Other processes are not taken into account since they play a minor role in typical AGN jets \cite{beall}.\\
The kinetic equation of electrons in the radiation zone becomes
\begin{align}
 \label{radzone}
	\partial_t N_{e^-} = &\partial_{\gamma}\left[\left(\beta_{s,e} \gamma^2 + P_{\text{IC}}(\gamma)\right) \cdot N_{e^-} \right] -\nonumber \\ & \frac{N_{e^-}}{t_{\text{rad,esc,e}}} + Q_{\text{pp}} + Q_{\text{p}\gamma -} + b^3\frac{n_{e^-}}{t_{\text{esc,e}}}
\end{align}
with $b=R_{\text{acc}}/R_{\text{rad}}$ and the pair production rate $Q_{\text{pp}}(\gamma)$ calculated from \cite{boettcherpair}. The production rate of $e^-$ from $\pi^-$ is calculated using \cite{kelneraharonian} eq. (30) with the corresponding $\Phi$-function of their paper. For positrons one finds
\begin{align}
 \label{radzone2}
	\partial_t N_{e^+} = &\partial_{\gamma}\left[\left(\beta_{s,e} \gamma^2 + P_{\text{IC}}(\gamma)\right) \cdot N_{e^+} \right] -\nonumber \\ & \frac{N_{e^+}}{t_{\text{rad,esc,e}}} + Q_{\text{pp}} + Q_{\text{p}\gamma +}
\end{align}
since there are no primary positrons in our model. Compton scattering losses of protons as well as $p + \gamma$ losses can be neglected following the discussion of \cite{disc1, disc2}, hence
\begin{align}
 \label{radzone3}
	\partial_t N_p = &\partial_{\gamma}\left[\left(\beta_{s,p} \gamma^2 \right) \cdot N_p \right] -\frac{N_p}{t_{\text{rad,esc,p}}} + b^3\frac{n_p}{t_{\text{esc,p}}}
\end{align}
The $Q_{\text{pp}}$, etc. gains and IC losses of course depend on the photon field which is calculated from the corresponding production and loss rates:
\begin{align}
 \label{radzone4}
	\partial_t N_{\gamma} & = R_s + R_c + R_{\pi^0} - c \left(\alpha_{\text{SSA}} + \alpha_{\text{pp}} \right) N_{\gamma} - \frac{N_{\gamma}}{t_{\text{ph,esc}}}
\end{align}
with the additional $\alpha_{\text{pp}}$ accounting for losses due to the $e^{\pm}$ pair-channel calculated from \cite{coppi}, eq. (4.3) and the $\pi^0$-decay channel as calculated from \cite{kelneraharonian}, eq. (11). The synchrotron production rate $R_s$ now consists of electron, positron and proton synchrotron radiation being produced due to eqn. \eqref{radzone} to \eqref{radzone3}, $R_c$ takes IC photons from $e^-$ and $e^+$ into account.\\
The model is tested with respect to energy and particle conservation of all contributing processes. Note that we do not introduce any ad hoc assumptions and all spectra arise selfconsistently during the time evolution of the model. This allows conclusions on the microphysics of the jet exploiting the variability of blazars as well, see \cite{weidinger02}.

\section{Results}
\begin{figure}[h]
  \vspace{5mm}
  \centering
  \includegraphics[width=3.in]{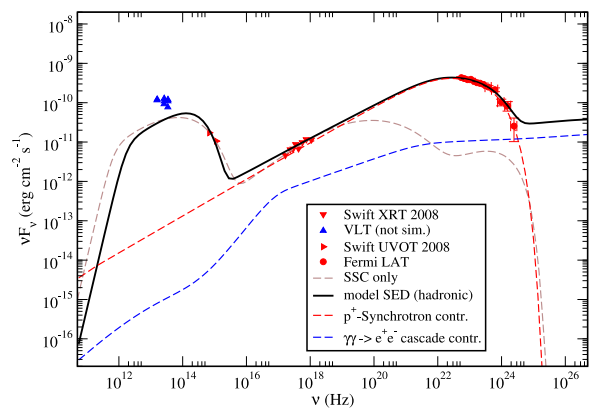}
  \caption{Modelled SED of \sourcesix~ including their individual processes, dashed red: proton synchrotron, dashed blue: subsequent cascade radiation. The grey line shows the modelling attempt using only $e^-$ with physically motivated parameters which fails to reproduce the VHE part. The data was taken from \cite{3C454_1}, the model parameters can be found in Table \ref{tab:lowstates}. The VLT observations are not simultaneous.}
  \label{fig:3C454_sed}
\end{figure}

The blazars used in the modelling summarized in Table \ref{tab:lowstates} show variability from the sub-hour timescale (\sourcethree) to outbursts lasting for days (\sourceone, \sourcefour) or at least have been observed in multiwavelength observation campaigns in different flux states (e.g. \sourcetwo). Table \ref{tab:lowstates} also shows the the model parameters found for each object for its lowstate emission as well as their blazar-type and redshift.
\begin{table*}[t]
\begin{center}
\begin{tabular}{l|cl||cccccccc}
\hline
\begin{footnotesize}Source\end{footnotesize} & \begin{footnotesize}$z$\end{footnotesize} & \begin{footnotesize}Type\end{footnotesize} & $Q_e (\text{cm}^{-3})$  & $Q_p (\text{cm}^{-3})$ & $B(\text{G})$ & $R_{\text{acc}}(\text{cm})$ & $R_{\text{rad}}(\text{cm})$ & $t_{\text{acc}}/t_{\text{esc}}$ & $a$ & $\delta$\\
\hline
\begin{footnotesize}\sourceone \end{footnotesize} & \begin{footnotesize}$0.034$\end{footnotesize} & \begin{footnotesize}HBL\end{footnotesize} & $3.1 \cdot 10^{5}$ & $0$ & $0.09$ & $2.2 \cdot 10^{14}$ & $5.0 \cdot 10^{15}$ & $1.20$ & $50$ & $34$\\
\begin{footnotesize}\sourcetwo\end{footnotesize} & \begin{footnotesize}$0.044$\end{footnotesize} & \begin{footnotesize}HBL\end{footnotesize} & $6.5 \cdot 10^{5}$ & $0$ & $0.10$ & $1.8 \cdot 10^{14}$ & $6.0 \cdot 10^{15}$ & $1.05$ & $50$ & $18$\\
\begin{footnotesize}\sourcethree\end{footnotesize} & \begin{footnotesize}$0.117$\end{footnotesize} & \begin{footnotesize}HBL\end{footnotesize} & $8.0\cdot 10^5$ & $0$ & $1.4$ & $1.0 \cdot 10^{13}$ & $5.0 \cdot 10^{14}$ & $1.13$ & $20$ & $49$\\
\begin{footnotesize}\sourcefour\end{footnotesize} & \begin{footnotesize}$0.182$\end{footnotesize} & \begin{footnotesize}HBL\end{footnotesize} & $6.3 \cdot 10^{4}$ & $0$ & $0.12$ & $6.0 \cdot 10^{14}$ & $3.0 \cdot 10^{15}$ & $1.11$ & $10$ & $44$\\
\begin{footnotesize}\sourcesix\end{footnotesize} & \begin{footnotesize}$0.859$\end{footnotesize} & \begin{footnotesize}FSRQ\end{footnotesize} & $3.8 \cdot 10^{7}$ & $4.2 \cdot 10^{8}$ & $10.2$ & $5.0 \cdot 10^{13}$ & $5.0 \cdot 10^{15}$ &$1.10$ & $5000$ & $43$\\
\hline
\end{tabular}
\caption{List of blazars used along with the parameters used for modelling the lowstate emissions with our model, see text for details.}\label{tab:lowstates}
\end{center}
\end{table*}
All parameters displayed in Table \ref{tab:lowstates} are in a physically sensible range compared to other observations like superluminal motion putting upper limits on $\delta$ or the observed variability timescales / Schwartzschlid radius lower boundaries on $R_{\text{rad}}$ and coincide within an order of magnitude with other modelling attempts by e.g. F. Tavecchio (\sourcetwo, see \cite{ruegamer}) while not relying on ad-hoc assumptions. The magnetic field for each blazar is high enough to confine the particles within the emission region, i.e. $r_{\text{gyr}} \approx \gamma m_i c/eB \ll R_{\text{rad}}$. Without introducing a significant number of non-thermal protons $Q_p$ in the case of \sourcesix~one fails to find parameters modelling the SEDs satisfying the mentioned criteria hence strongly disfavoring a purely SSC scenario for FSRQ like objects.\\
With our timedependent code we are in the unique position to exploit the variability of blazars from HBLs to FSRQs providing information about the nature of the physics within the jet. This can be used to i) constrain the model parameters more tightly allowing together with the selfconsistency of the model microphysical interpretations of them and ii) to discriminate between purely leptonic jets and hadronic scenarios since each leaves a typical imprint in lightcurves.\\
For the HBLs \sourcefour~ and \sourcethree~ the timedependent results can be found in \cite{weidinger01} and \cite{weidinger02} respectively. The lightcurves in the VHEs can be traced remarkably well simply by injecting more primary electrons into the blob.
\begin{figure}[h]
  \vspace{5mm}
  \centering
  \includegraphics[width=3.in]{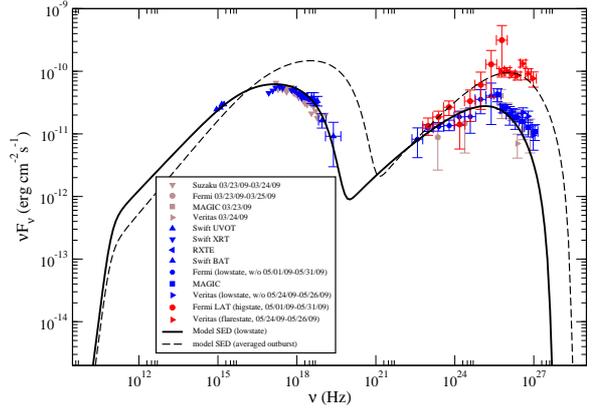}
  \caption{Modelled lowstate emission of \sourceone~ (solid line, parameters see Table \ref{tab:lowstates}) and the averaged outburst in 05/2009 (dashed line, averaging and parameters see text). The lowstate was found taking the multiwavelength observations of \cite{mkn501fermi} as well as \cite{mkn50102} (brown points) into account. The VHE data has been EBL corrected, see \cite{mkn501fermi}.}
  \label{fig:mkn501_sed}
\end{figure}
In the case of \sourceone, as well being modelled purely leptonic, the variability seems to have a different nature which can clearly be seen in Fig. \ref{fig:mkn501_sed}. Unlike \sourcefour~ or \sourcethree~ the peak in the VHEs shifts towards higher energies during the outburst observed by VERITAS and Fermi in May 2009. Similar behaviour was, even more prominently, observed by BeppoSAX during the famous outburst 1997 in the X-Rays \cite{mkn5011997} making the scenario somehow persistent even when considering two unrelated events (blobs). The recent flare was modelled as a linear dropoff in the magnetic field from $B = 0.09~\text{G}$ to $B = 0.043~\text{G}$ over a time of $t_{\text{drop}} = 11~\text{d}$ which consistently is accompanied by a dropoff in the injection of particles to $Q_e = 1.60 \cdot 10^5~\text{cm}^{-3}$ and of $t_{\text{acc}}/t_{\text{rad}}$ to $1.0$. After another $t_{\text{high}} = 40~\text{d}$ these values linearly took their lowstate values displayed in Tab. \ref{tab:lowstates} within $t_{\text{rise}} = 28~\text{d}$. The average over the whole outburst results in the highstate (dashed line in Fig. \ref{fig:mkn501_sed}) of \sourceone, all times are given in the observer's frame.
\begin{figure}[h]
  \vspace{5mm}
  \centering
  \includegraphics[width=3.in]{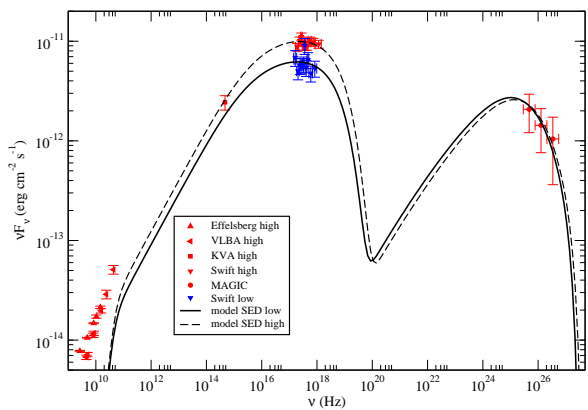}
  \caption{Modelled SED of \sourcetwo~ in two different flux states (lowstate: solid line, parameters see Table \ref{tab:lowstates}, highstate: dashed line, parameters see text). The data was was almost simultaneously taken, see \cite{ruegamer} for details.}
  \label{fig:2344_sed}
\end{figure}
The highstate of \sourcetwo~ was found by setting $\delta = 23$ and $Q_e = 3.75 \cdot 10^5~\text{cm}^{-3}$ instead of the values found in Tab. \ref{tab:lowstates} suggesting observing radiation from two different blobs, therefore no time evolution was tested at this particular HBL, although a speed-up or altering angle to the line of sight of the emitting region might be possible but unlikely interpretations.\\
Concerning \sourcesix~ a significant number of protons had to be injected into the blob which are accelerated due to Fermi I and Fermi II processes up to $10^9~\text{GeV}$ for a small fraction of them. Purely leptonic fits lead to arbitraryly low magnetic fields as well as $\delta \gg 100$ and thus were neglected. The radiation in the optical regime essentially is synchrotron emission from primary electrons, in the X-Rays the SED is dominated by proton synchrotron and reprocessed radiation (synchrotron radiation of $e^{\pm}$ pairs cascading down from pion production and their radiation), see Fig. \ref{fig:3C454_sed}.
In the Fermi-LAT energy range the SED of \sourcesix~ seems to be dominated by proton synchrotron emission, see Fig. \ref{fig:3C454_sed}. Concerning variability the proton synchrotron radiation in the X-rays is providing a very stable flux-basis with the proton synchrotron timescale of the order of months. Shorter timescales as observed during the flares in December 2009 \cite{3C454_2} are hence due to cascade radiation being enhanced by fluctuations in the primary electrons and/or their synchrotron emission itself due to $B$ field changes (the latter might explain the observed flaring behaviour of \cite{3C454_2}). In both cases the cascade radiation is outmatching the proton synchrotron radiation of the lowstate while cascading down.
\section{Conclusion and Discussion}
We were able to model all four low-redshift blazars considered with our selfconsistent model without introducing non-thermal protons within the jet. Making use of the timedependence of our model we constrained the parameters down to the values found in Table \ref{tab:lowstates}. However the variability of HBLs seems to have no common cause. In the case of \sourcethree~ and \sourcefour~ injecting more particles into the blob (i.e. density fluctuations within the jet, as the blob travels through them) reproduces the high states and lightcurves. Concerning \sourceone~ the variability seems to have a different nature: a dropoff in the magnetic field accompanied with a consistent dropoff in the particle density. This sort of outburst was observed in the very recent flare we modelled and also back in 1997 by BeppoSAX in the x-rays. This might be related to the stability of the jet. The variability of \sourcetwo~ suggests observing radiation from two different blobs, see Sect. 3 for details. Nevertheless all four HBLs allow for large amplitute, short time variability and can be observed at very different flux states over the years as e.g. \sourceone.
The situation is different for the rather distant FSRQ \sourcesix. The SED in the Fermi energy range seems to be dominated by proton synchrotron radiation which has a variability timescale of the order of days as observed in outbursts \cite{3C454_2, 3C454_3}. Short time X-ray variability is explained by cascade radiation dominating over the prominent lowstate proton synchrotron emission during an outburst. The parameters found for \sourcesix~ agree well with other hadronic models of by M. B\"ottcher \cite{boettcher3C279} for other FSRQs, e.g. \sourcefive.\\
Concluding HBLs do not allow hadrons to be accelerated to energies above thermal due to their relatively low magnetic field being not able to confine them within the emitting region for a significant time. Hence FSRQs seem to be the most promising candidates among blazars as sources of extragalactic cosmic rays. However, their lower peak energies compared to HBLs point towards maximum proton energies (after accounting for the blob's bulk motion) above the ankle in the spectrum of CRs but not to energies of UHECRs. \textit{Acknowledgements}
MW thanks Elitenetzwerk Bayern and GK1147 for their support. FS receives support from DFG through grant SP 1124/1. 

%

\begin{scriptsize}

\end{scriptsize}
\clearpage

\end{document}